\documentclass{JHEP3}

\usepackage{amsfonts}
\usepackage{amssymb}
\usepackage{amscd}
\usepackage{latexsym}
\usepackage{amsbsy}

\def\a{\alpha}
\def\b{\beta}
\newcommand{\g}{\gamma}
\def\d{\delta}
\newcommand{\D}{\Delta} 

\newcommand{\ep}{\epsilon}

\newcommand{\la}{\lambda}
\newcommand{\La}{\Lambda}

\newcommand{\si}{\sigma}

\def\cH{\mbox{${\cal H}$}}

\def\cM{\mbox{${\cal M}$}}

\def\dim{{\rm dim}}
\def\diag{{\rm diag}}
\def\det{{\rm det}}

\def\tr{{\rm tr}}

\def\exp{{\rm exp}}

\def\llap#1{\hbox to 0pt{\hss#1}}
\def\pola{a\llap{\hbox{\char'30\kern-1.2pt}}}
\def\pole{e\llap{\hbox{\char'30\kern-.8pt}}}

\newcommand{\non}{\nonumber\\}

\def\half{{\mbox{\small  $\frac{1}{2}$}}}

\newcommand{\beq}{\begin{equation}}
\newcommand{\eeq}{\end{equation}}
\newcommand{\beqa}{\begin{eqnarray}}
\newcommand{\eeqa}{\end{eqnarray}}
\newcommand{\barr}{\begin{array}}
\newcommand{\earr}{\end{array}}
\newcommand{\ben}{\begin{enumerate}}
\newcommand{\een}{\end{enumerate}}
\newcommand{\bit}{\begin{itemize}}
\newcommand{\eit}{\end{itemize}}

\newcommand{\refeq}[1]{(\ref{#1})}

\def\C{{\mathbb C}}
\def\N{{\mathbb N}}
\def\Z{{\mathbb Z}}

\def\gcheck{g^{\vee}}
\def\obar{\overline}

\def\trr{\triangleright}
\def\rep{representation }
\def\reps{representations }

\def\cC{{\cal C}}
\def\cF{{\cal F}}
\def\cG{{\cal G}}

\def\cM{{\cal M}}
\def\cR{{\cal R}}

\def\cG{{\cal G}}
\def\cU{{\cal U}}

\def\nn{\nonumber}

\def\tens{\otimes}

\def\mg {{\mathfrak g}}
\def\mgh {\widehat{\mathfrak g}}
\def\mk {{\mathfrak k}}
\def\GcG{\cG_L \tens^\cR \cG_R}
\def\gcg{U_q(\mg_L \times \mg_R)_\cR}
\def\eps{\varepsilon}

\def\cH {{\cal H}}

\def\Mo{M^{(0)}}

\newcommand{\matR}[4]{R{}^{#1}{}_{#2}\,{}^{#3}{}_{#4}\,}

\def\rara{\rangle\rangle}

\title{A quantum algebraic description of D-branes on group
  manifolds~{}} 

\author{      
Jacek Pawe\l czyk \\
 Institute of Theoretical Physics\\ Warsaw University, Ho\.{z}a 69,
PL-00-681 Warsaw, Poland\\ \email{Jacek.Pawelczyk@fuw.edu.pl}
}
\author{Harold Steinacker
\\ 
Sektion Physik der Ludwig--Maximilians--Universit\"at M\"unchen\\
Theresienstr.\ 37, D-80333 M\"unchen  \\
\email{Harold.Steinacker@physik.uni-muenchen.de}
}

\abstract{We propose an algebraic description 
of (untwisted) D-branes on compact group manifolds $G$ using
quantum algebras related to $U_q(\mg)$. It 
reproduces the
known characteristics of stable branes in the 
WZW models, in particular their configurations in $G$, energies
as well as the set of harmonics. Both generic and
degenerate branes are covered. \\[1ex]

 }


\preprint{LMU--TPW 03/02\\IFT--02/09\\hep-th/0203110}

\begin{document}

\newpage

\section{Introduction}

Recently the structure of $D$-branes in a $B$ field background 
has attracted much attention. The case of flat branes in a 
constant $B$ background has been studied extensively 
(see e.g. \cite{douglas}), and leads to quantum spaces
with a Moyal-Weyl star product. This was later generalized to
non-constant, closed $B$  \cite{schupp2}. 
A rather different situation is given by $D$-branes
on compact Lie groups $G$, which carry a (NSNS) $B$ field which is
not closed.  
It has been shown, using CFT \cite{as} and  DBI (Dirac-Born-Infeld)
\cite{jbds} descriptions that stable branes can 
wrap certain conjugacy classes in the group manifold. 
On the other hand, the matrix model \cite{myers} and CFT calculations 
\cite{ars2} led to a beautiful picture where, in a 
special limit, the macroscopic branes are formed as a
bound state of $D0$-branes.
Attempting to unify 
these various approaches, we proposed
in a recent paper \cite{PS} a matrix description of $D$ - branes on
$SU(2)$.
This led to a quantum algebra based on quantum group symmetries,
which reproduced all static properties of stable
$D$-branes on $SU(2)$.

In the present paper, we generalize the methods of \cite{PS} 
and propose a simple and compact description 
of all (untwisted) D-branes on group manifolds $G$, using
quantum algebras related to $U_q(\mg)$. 
More specifically, we show that a simple algebra 
known for more than 10 years as reflection equation (RE)
leads to precisely the same 
branes as the DBI approach or the WZW model. 
It not only reproduces their configurations in $G$, 
i.e. the positions of the corresponding conjugacy classes, 
but also a (quantized) algebra of functions on the branes which turns
out to be essentially the same as 
given by CFT. 
Moreover, both generic and degenerate branes are predicted,
again in agreement with the CFT results. In particular, 
we identify branes on $SU(N+1)$ which are quantizations of $\C P^N$,
and we show that they precisely correspond to the fuzzy $\C P^N$ 
constructed in \cite{bala,stroh}. 

We do not attempt here to recover all known branes on $G$, such as
twisted branes or ``type B branes'' \cite{mms}
but concentrate on the untwisted branes.
Given the success and simplicity
of our description, it seems quite possible, however, that these other
branes
are described by RE as well. 
Our results can be briefly summarized as follows:
$D$-branes on $G$ are described by the RE. A large class of irreps
of RE corresponds to irreps of $U_q(g)$, and describes
untwisted branes.  

We should point out that all mathematical constructions are basically
well-known. In spite of this,
we tried to make the paper accessible to a wide audience, by 
giving the basic constructions and results in the main body of the paper
while postponing many technical aspects to the Appendix. 
The paper can be read from a variety of viewpoints, starting from 
a string theorists perspective emphasizing the agreement
with other approaches, but also from a more algebraic
point of view given the simple and compact description of quantized
adjoint orbits on $G$.

The paper is summarized as follows. Some basic facts about
(untwisted) 
$D$-branes on compact Lie groups and their description in CFT
are recalled in Section 2, with emphasis on those aspects which are
useful in later considerations. We argue that the finite 
set of primaries of BCFT of a D-branes can be interpreted in terms
of NCG, i.e. they provide 
a picture of D-branes as quantum manifolds. Moreover, we
claim that the appropriate symmetry algebra is a particular
quantum group, replacing (in a
sense) the chiral affine algebra $\mgh_L\times\mgh_R$. 
The quantum manifold is defined as
an associative algebra, generated by some elements subject to certain
relations. In Section 3 we  
postulate these relations (the so-called reflection equations RE),
and discuss their basic properties.
In particular, we show that RE has all the required properties under 
the quantum group. The D-branes are then obtained 
from representations of RE. These are
are studied in Section 4, where we justify our claims.
In particular, we calculate the positions of the D-branes
on the group manifold, their energies,
and perform the harmonic analysis on their quantized world-volumes. 
Our constructions are illustrated
in some examples in Section 5, studying fuzzy $\C P_q^N$ in more 
detail, and reviewing the $SU(2)$ case in order to make clear
the connection with our previous paper \cite{PS}.
Finally, some technical discussions are collected
in the Appendices.

\section{CFT and classical description of untwisted D-branes}
\label{cft}
This section deals with the CFT description of branes in 
WZW models on $G$, and their
classical interpretation as certain sub-manifolds in the group manifold $G$. 
All the results presented here are well known and serve only as inspiration to
the algebraic considerations in the rest of the paper. The reader who is not
familiar with CFT and string theory may skip this part of the paper and go
directly to the Subsection \ref{sec:c-branes}.

\subsection{Some Lie algebra notations}

We collect some notations used throughout this paper.
$\mg$ denotes the (simple, finite-dimensional) Lie algebra of $G$, with 
Cartan matrix $A_{ij}=2\frac{\a_i\cdot\a_j}{\a_j\cdot\a_j}$.
Here $\cdot$ is the
Killing form which is defined for arbitrary weights, and 
$\a_i$ are the simple roots.
The set of dominant integral weights is denoted by
\beq
P^+ = \{\sum n_i \La_i;\;\; n_i \in \Z_{\geq 0} \},
\eeq 
where the fundamental weights $\La_i$ satisfy
$\a_i\cdot\La_j = d_{\a_i} \d_{ij}$,
and the length of a root $\a$ is $d_\a = \frac{\a \cdot \a}2$.
The Weyl vector is the sum over all positive roots, 
$\rho = \frac 12 \sum_{\a>0} \a$.
For a positive integer $k$, 
one defines the ``fundamental alcove'' in weight space as
\beq
P_k^+ = 
 \{\la \in P^+; \; \la\cdot\theta \leq k\}
\label{fundalcove}
\eeq
where $\theta$ is the highest root.
It is a finite set of dominant integral weights.
For $G=SU(N)$, this is explicitly
$P_k^+ = \{\sum n_i \La_i;\;\sum_i n_i \leq k\}$. 
We shall normalize the Killing form such that
$d_\theta = 1$, so that the dual Coxeter number is given by
$\gcheck = (\rho + \frac 12 \theta)\cdot \theta$,
which is $N$ for $SU(N)$.

For any weight $\la$, we define $H_\la \in \mg$ 
to be the Cartan element which takes the value
$H_\la v_\mu =(\la\cdot\mu)\;v_\mu$
on vectors $v_{\mu}$ with weight $\mu$ in some representation.
We shall consider only finite-dimensional representations 
(=modules) of $\mg$.
$V_{\la}$ denotes the irreducible
highest-weight module of $G$ with highest weight $\la \in P^+$, and
$V_{\la^+}$ is the conjugate (=dual) module of $V_{\la}$.
The defining \rep of the classical matrix
groups $SU(N)$, $SO(N)$, and 
$Sp(N)$ will be denoted by $V_N$, being $N$-dimensional.

\subsection{WZW D-branes}
\label{sec:wzw-b}

The WZW model is specified by a group $G$ and a level $k$
\cite{b-fuchs,b-diF}. We shall consider only simple, compact groups ($G$ will
be $SU(N)$ mainly),
so that the level $k$ must be a positive integer. 
The WZW branes can  be
described by boundary states $\vert B\rara\in \cH^{\rm closed}$ respecting  a
set of boundary conditions. 
A large class of boundary conditions 
is of the form
\beq
 \biggl( J_{n} + \tilde\g(\tilde J_{-n}) \biggr) \vert
B \rangle \rangle =0 \qquad n\in {\bf Z} 
\label{twinbc}
\eeq
where $\tilde\g$ is an auto-morphism of the affine Lie algebra 
$\mgh$\footnote{
$\mg$ is the horizontal algebra of $\mgh$, and the Lie algebra of $G$.}.
Here  $J_n$ are the modes of the left-moving currents and $\tilde
J_n$ are the modes of the right-moving currents.  Boundary states
with $\tilde\g=1$ 
are called ``symmetry-preserving branes'' or "untwisted branes":
these are the object of interest in this  paper.
The untwisted ($\tilde\g=1$) 
boundary condition \refeq{twinbc} breaks half of the
symmetries of the WZW model $\mgh_L\times\mgh_R$ down to the vector part
$\mgh_V$.

The condition \refeq{twinbc} alone does not define a good boundary state:
one must also impose open-closed string duality of the 
amplitude describing
interactions of branes. This leads to so called
Cardy (boundary) states. 
For the untwisted case they are labelled by
$\lambda\in P_k^+$ corresponding to integrable irreps of  
$\widehat \mg$, which are
precisely the weights in the ``fundamental alcove'' \refeq{fundalcove}.
Therefore the untwisted branes are in one-to-one correspondence with 
$\lambda\in P_k^+$.
The CFT description yields also an important formula for the energy 
of the brane $\la$, 
\beq\label{crdymss}{
E_\la=\prod_{\a > 0}
{\sin\biggl(\pi {\alpha\cdot (\lambda+ \rho)\over k+\gcheck}\biggr)
\over \sin\biggl(\pi {\alpha\cdot  \rho\over k+\gcheck}\biggr)}
}\eeq
For $k\gg N $, one can expand the denominator in \refeq{crdymss} to obtain a
formula which compared with
DBI \cite{mms} shows that the leading $k$-dependence
fits perfectly with the interpretation of a brane wrapping 
once a conjugacy class given by an element $t_\la$ of the maximal torus of $G$
(see the next subsection). 

The CFT provides hints towards the 
description of branes as quantum manifolds. 
It is known that the dynamics of D-branes is
 given by open string excitations. The relevant operators, entering as
 building blocks of the string operators, are the primary fields of the BCFT
 with the symmetry algebra of the unbroken part $\mgh_L\times\mgh_R$,
i.e. $\mgh_V$. The number of lowest conformal weight primaries is 
finite for any compact WZW model (in general for any RCFT).
In the $k\to\infty$ limit, the primaries can be
interpreted as corresponding to a (finite dimensional) algebra of
functions on the brane (see \cite{ars,FFFS} and Section 
\ref{subsec:harmonics}). 
For finite $k$, the interpretation is
not that clear because the candidate algebra 
as given in \cite{ars} is not associative. 
However as explained in \cite{ars,PS} for $\mg=su(2)$, 
the algebra becomes associative 
after "twisting" (resulting in a modification of the
product of the primary fields),
so that it can be considered as an algebra of functions of a quantum
manifold. Then the primaries become modules of the quantum group
$U_q(su(2))$. We 
argued in \cite{PS} that the relations defining the algebra of
functions on the quantum manifold is invariant under the full chiral
counterpart of the chiral current algebra, i.e. under 
$U_q(su(2)_L \times su(2)_R)_\cR$.

Here we shall follow
the line of reasoning of \cite{PS} replacing $\mg=su(2)$ by any compact,
simple Lie algebra $\mg$,
noting that the technical arguments for twisting and associativity
generalize. We shall therefore assume that one can modify 
the product of primary fields such that they form an 
associative algebra, 
and transform under a suitable quantum group $\gcg$ (or $\GcG$)
as given below.

\subsection{The classical description of $D$--branes on group manifolds.}
\label{sec:c-branes}

The D-branes whose quantum description has been given in the 
previous subsection have a nice geometrical interpretation: 
they correspond to the conjugacy classes 
of the group manifold under the adjoint action. Here we describe 
some properties of
those sub-manifolds. The results presented in the forthcoming sections can 
also be viewed as a quantization of those sub-manifolds.

Let $G$ be the classical group manifold (we will consider mainly $SU(N)$,
but all constructions can be used for other groups 
such as $SO(N)$, $USp(N)$ as well).
At the classical level, the $D$--branes under consideration are described 
by (twisted) conjugacy classes of the form 
\beq
\cC(t) = \{g t \g(g)^{-1}; \quad g \in G\}.
\label{conj-classes}
\eeq
Here $\g$ is an auto-morphism of $G$, which is 
related to that of \refeq{twinbc}. In this paper we shall consider only  
trivial $\g$, leaving the $\g\neq id$ case to a future publication.
One can take $t$ belonging to a maximal torus $T$ of $G$, i.e. 
$t$ is a diagonal matrix for $G = SU(N)$. 
Then $\cC(t)$ can be viewed as homogeneous spaces 
(see Appendix \ref{subsec:harmonic-brane}):
\beq
\cC(t) \cong G/K_t.
\label{coset}
\eeq
Here $K_t = \{g \in G:\; [g,t] = 0\}$ is the stabilizer of $t \in T$.
``Regular'' conjugacy classes are those with $K_t = T$, 
and they are isomorphic to $G / T$. In particular, their dimension is 
$dim(\cC(t)) = dim(G) - rank(G)$.
``Degenerate'' conjugacy classes have a larger stability group $K_t$,
hence their dimension is smaller;
e.g. at the extremal case $\cC(t=1)$ is a point.
These conjugacy classes are invariant under the adjoint action
\beq\label{adj-action}
G_V^{-1} \cC(t) G_V = \cC(t)
\eeq
of the vector subgroup 
$G_V \hookrightarrow G_L \times G_R$,
which is diagonally embedded in the group of (left and right) motions
on $G$. This reflects the breaking 
$\mgh_L\times\mgh_R \to \hat\mg_V$.
We want to preserve this symmetry pattern
in the quantum case, in a suitable sense.

\paragraph{The space of harmonics on $\cC(t)$.}

A lot of information about the spaces $\cC(t)$ can be obtained from the 
harmonic analysis, i.e. by decomposing scalar fields on $\cC(t)$
into harmonics under the action of the (vector) symmetry $G_V$. 
This is particularly useful here, because 
quantized spaces are described in terms of their
algebra of functions. The decomposition of this 
space of functions $\cF(\cC(t))$ into harmonics can be calculated explicitly
using \refeq{coset}, and it must 
be preserved after quantization,
at least up to some cutoff.
Otherwise, the quantization would not be admissible.
One finds (see Appendix \ref{subsec:harmonic-brane} and \cite{FFFS}) 
\beq
\cF(\cC(t)) \cong  \bigoplus_{\la \in P^+} \; mult_{\la^+}^{(K_t)}\; V_{\la}.
\label{FCt-decomp}
\eeq
Here $\la$ runs over all dominant integral weights $P^+$,
$V_{\la}$ is the corresponding highest-weight G-module, and
$mult_{\la^+}^{(K_t)}$ is the dimension of the subspace of 
$V_{\la^+}$ which is invariant under $K_t$.

\paragraph{Characterization of the stable $D$--branes.}

From the CFT \cite{as,FFFS} and DBI  
considerations \cite{jbds,mms},
one finds that there is only a finite
set of stable $D$--branes on $G$ (up to global motions),
one for each integral weight $\la\in P_k^+$.
They are given by $\cC(t_{\la})$ for
\beq
t_{\la} = \exp(2 \pi i \frac{H_\la + H_\rho}{k + \gcheck}).
\label{stable-branes}
\eeq
The restriction to $\la\in P_k^+$ follows from the fact that
in general, different integral $\la$ may label the same conjugacy class.
Because the exponential in \refeq{stable-branes} is periodic, this 
happens precisely if
the weights are related by the affine Weyl group, which is 
generated by the ordinary Weyl group together with translations of the form
$\la \rightarrow \la + (k+\gcheck) \frac{2\a_i}{\a_i\cdot \a_i}$.
Hence one should restrict the weights to 
the fundamental domain of this affine Weyl group, which is the 
fundamental alcove $P_k^+$ \refeq{fundalcove} but with $k \to k + \gcheck$.

Information about the location
of these (untwisted) branes in $G$
is provided by the quantities 
\beq
s_n = \tr(g^n) = \tr(t^n), \quad g \in \cC(t)
\eeq
which are invariant under the adjoint action \refeq{adj-action}.
The trace is over the defining \rep $V_N$ 
($=V_{\La_1}$ in the case of $SU(N)$, 
where $\La_1$ is the fundamental weight) of the matrix group $G$,
of dimension $N$. For the classes  $\cC(t_{\la})$,
they can be easily calculated:
\beq
s_n =  \tr_{V_N}\; (q^{2n (H_\rho +H_\la)}) = 
   \;\sum_{\nu \in V_N}\; 
   e^{2 \pi i  n \frac{(\rho +\la)\cdot \nu}{k+\gcheck}}
\label{sn}
\eeq
where 
$$
q = e^{\frac{i \pi}{k+\gcheck}}.
$$
The $s_n$ are independent functions 
of the weight $\la$ for all $n = 1,2,..., rank(G)$,   which
completely characterize the class $\cC(t_{\la})$.
These functions have the great advantage that their quantum analogs 
(\ref{central2-n}) can be calculated exactly.

An equivalent characterization of these conjugacy classes is provided by 
a characteristic equation: for any $g \in \cC(t_\la)$, the relation 
$
P_\la(g) = 0
$
holds in $Mat(V_N,\C)$, 
where $P_\la$ is the polynomial 
\beq
P_\la(x) = \prod_{\nu \in V_N} (x-q^{2(\la+\rho)\cdot\nu}).
\label{charpoly-class}
\eeq
This follows immediately from (\ref{stable-branes}):
$t_\la$ has the eigenvalues $q^{2(\la+\rho)\cdot\nu}$
on the weights $\nu$ of the defining \rep $V_N$.
Again, we will find analogous characteristic equations in the 
quantum case.

\section{Quantum algebras and symmetries for branes}
\label{sec:LR-symm}

We expect that the relevant quantum spaces are described by quantum
algebras $\cM$ which transform appropriately under a quantum symmetry.
To find $\cM$ we shall make an ``educated guess'' based on the considerations
in Section \ref{sec:wzw-b}, and justify it  
by comparing its predictions with the results listed above. 
Thus first  we postulate the form of the relations
between generators of the quantum algebra. We expect the relations to be
at most quadratic in generators, and to have appropriate covariance under the
action of a quantum group which should correspond to the 
chiral $\mgh_L\times\mgh_R$. This quantum group will be $\gcg$.
Moreover we require the central terms of the algebra to be invariant under the
"vector" subalgebra of this quantum symmetry. Thus our constructions mimic the
symmetry pattern and its breaking by the D-branes in CFT.  
This is discussed in Section \ref{sec:qsym}.

\subsection{The module algebra}

The discussion invoked in the end of Section \ref{sec:wzw-b} 
suggests that $\cM$ should be a module 
algebra\footnote{see  Appendix \ref{a:symmetries} for the 
mathematical definition} under some quantum group. 
Moreover, it suggests that this quantum group is a version of 
$U_q(\mg)$, the \reps of which are parallel to those of 
$\mgh$ of the WZW model. Since we are considering matrix groups $G$, 
we assume that the appropriate quantum (module) algebra $\cM$ is generated by 
elements $M^i_j$ with indices $i,j$ in the defining \rep $V_N$ of $G$,
subject to some commutation relations and constraints.
With hindsight, we claim that these relations are given by
the so-called reflection
equation (RE) \cite{re}, which in a short notation  reads
\beq
R_{21} M_1 R_{12} M_2 = M_2 R_{21} M_1 R_{12}.
\label{re}
\eeq
Here $R$ is the $\cR$ matrix  of 
$U_q(\mg)$ in the defining representation. 
Displaying the indices explicitly, this means
\beq
({\rm RE })^{\ i\ k}_{\ j\ l}: \qquad
\matR kaib\ M^b_c\ \matR cjad\ M^d_l = M^k_a\ \matR abic\ M^c_d\ \matR djbl.
\label{re-i}
\eeq
The indices $\{i,j\}$, $\{k,l\}$ correspond to the first (1) and the
second (2) vectors space $V_N$ in \refeq{re}. 
Some examples of algebras generated by RE relations are presented 
in Section \ref{sec:examples}.
For $q=1$, this reduces to $[M^i_j, M^k_l] = 0$.
Because $\cM$ should describe the quantized group
manifold $G$, we need to impose constraints which ensure that the
branes are indeed embedded in such a quantum group manifold. 
In the case $G=SU(N)$, these are
$\det_q(M)=1$ where $\det_q$ is the so-called quantum 
determinant \refeq{qdet}, and
suitable reality conditions imposed on the generators
$M^i_j$. Both will be discussed below.

Following \cite{PS}, the $M^i_j$'s can also be thought of as some matrices 
(as in Myers model \cite{myers}) 
out of which we can form an action invariant under
the relevant quantum groups. The action has the structure 
$S=\tr_q(1+...)$, where dots represent some 
expressions in the $M$'s
(the quantum trace is defined in \refeq{qdim}).
The point of \cite{PS} was that  for some
equations of motion, the "dots"- terms vanish
on classical configurations. 
We postulate that the equations of motion for $M$
are given by RE \refeq{re}.
If so, then their energy is equal to
\beq\label{energy}
E=\tr_q(1).
\eeq
As we shall see this energy is not just a constant (as might be suggested by
the notation), but it depends on the representations of the 
algebra, where it becomes the quantum dimension \refeq{qdim}. 

We should mention here that RE 
appeared more then 10 years ago in the context of the boundary integrable
models, and is sometimes called boundary YBE \cite{re}.
Hence one might also think of
\refeq{re} as being analogs of the boundary condition \refeq{twinbc}.  
As we shall see, RE has indeed similar symmetry properties. 
This is the subject of the the following subsection.

\subsection{Quantum symmetries of RE}\label{sec:qsym}

Since $\cM$ is supposed to be a module algebra, we have to
specify under which quantum group it transforms. 
The construction of the quantum symmetry algebra is a 
straightforward generalization of  the approach of \cite{PS}, 
replacing $su(2)$ by $\mg$. However we 
found it more convenient to work with a dual version of this symmetry,
which leads directly to the desired results. 
We shall present here a simple practical version and postpone
the precise mathematical definitions to Appendix \ref{a:symmetries}.

There are 2 equivalent ways to look at the symmetry of RE, involving
the Hopf algebras
$\cG_L \tens^\cR \cG_R$ and $U_q(\mg_L \times \mg_R)_\cR$ respectively, 
which are dual to each other. 
We first assume that the matrix $M$ transforms as
\beq\label{st-coaction}
M^i_j \to (s^{-1}M t)^i_j
\eeq
where $s^i_j$ and $t^i_j$ generate the algebras $\cG_L$ and 
$\cG_R$ respectively, which both coincide with the 
well--known quantum groups $Fun_q(G)$ as defined in \cite{FRT} so e.g.
$
s_2s_1 R =R s_1s_2 ,\ t_2t_1 R=R  t_1t_2.
$\footnote{$R$ with suppressed indices means $R_{12}$.}
In  \refeq{st-coaction} matrix multiplication is understood.
This is a symmetry of RE if we impose that 
(the matrix elements of) $s$ and $t$ commute with $M$, and additionally satisfy
$ s_2 t_1  R= R t_1s_2$. Notice that \refeq{st-coaction} is a quantum analog
of the action of the classical isometry group $G_L\times G_R$ on classical
group element $g$ as in Section (\ref{sec:c-branes}). 

Symmetries become powerful only because they have a group-like
structure,
i.e. they can be iterated. In the above language this means that 
we can define a  Hopf algebra (called from now on $\GcG$):
\beqa\label{gg-algebra}
&&s_2s_1\ R =R\  s_1s_2 , \quad  t_2t_1\  R=R\  t_1t_2 , \quad s_2 t_1\ 
R= R\
t_1s_2 \\ 
&&\D s=s\otimes s,\quad \Delta t=t\otimes t,\\
&&S(s)=s^{-1},\quad \ep(s^i_j)=\d^i_j,\quad S(t)=t^{-1},\quad 
\ep(t^i_j)=\d^i_j
\eeqa
(here $S$ is the antipode, and $\ep$ the counit).
The inverse matrices $s^{-1}$ and $t^{-1}$ are defined after 
suitable further (determinant-like) constraints on $s$ and $t$ are
imposed,
as in \cite{FRT}.
Formally, $\cM$ is a right 
$\cG_L \tens^\cR \cG_R$ - comodule algebra;
see Appendix \ref{a:symmetries} for further details.

Furthermore, $\cG_L \tens^\cR \cG_R$ can be mapped to a 
vector Hopf algebra $\cG_V$ with
generators $r$, by 
$s^i_j\otimes 1\to r^i_j$ and $1\otimes t^i_j\to r^i_j$ (thus basically 
identifying $s=t=r$ on the rhs). 
The (co)action of $\cG_V$ on the $M$'s is then
\beq\label{v-coaction}
M^i_j \to (r^{-1}M r)^i_j.
\eeq

Equivalently, we can consider the Hopf algebra
$U_q(\mg_L \times \mg_R)_\cR$  which is dual to 
$\cG_L \tens^\cR \cG_R$. For the details we refer to 
Appendix \ref{a:symmetries}; 
we only state here that it is generated by 2 copies 
$U_q(\mg_L)$, $U_q(\mg_L)$ of $U_q(\mg)$,
which act on the the generators of $\cM$ as
\beq
(u_L \otimes u_R) \trr M^i_j = \pi^i_k(Su_L) M^k_l \pi^l_j(u_R)
\label{UU-action}
\eeq
where $\pi()$ is the defining \rep $V_N$ of $U_q(\mg)$. 
This is a symmetry of $\cM$ 
in the usual sense, because the rhs is again an element in $\cM$.
The ``vector" part of this symmetry
is obtained using the Hopf algebra map
$u \in U_q(\mg_V) \to \Delta(u) \in U_q(\mg_L \times \mg_R)_\cR$. It
acts on $\cM$ as
\beq\label{q-adj}
 u \trr M^i_j = \pi^i_k(S u_1) M^k_l \pi^l_j(u_2)
\eeq
where $u_1\otimes u_2=\Delta(u)$ is the standard coproduct
of  $U_q(\mg)$. 

We would like to stress here two crucial points in our construction:
the first is the existence of 
a vector sub-algebra $U_q(\mg_V)$ of $\gcg$ (and the analogous
notion for the dual $\GcG$). This is important because the
central terms of $\cM$ which characterize its \reps will be
invariant only
with respect to that $U_q(\mg_V)$ (and $\cG_V$). This will allow to interpret
these sub-algebras as isometries of the quantum D-branes. 
The second point is the fact that the RE imposes very similar 
conditions on the symmetries and their breaking as the original BCFT WZW
model described in section \ref{sec:wzw-b} does.

\subsection{Central elements of RE}
\label{sec:cen}

Below we discuss some general properties of the algebra defined by
\refeq{re}. We need to find the central elements,
which are expected to characterize its irreps. This problem was
solved in the second paper of \cite{kss}.
The (generic) central elements of the algebra \refeq{re} are 
\beq\label{central2-n}
c_n=\tr_q(M^n)\equiv \tr_{V_N}(M^n\ v) \; \in\cM,
\eeq
where the trace is taken over the 
defining representation $V_N$, and
\beq
v = \pi(q^{-2H_\rho})
\eeq
is a numerical matrix which satisfies $S^2(r)=v^{-1} r v$
for the generator $r$ of $\cG_V$.
These elements $c_n$ are independent for $n = 1, 2,..., rank(G)$.
A proof of centrality can be found e.g. in the book \cite{majid},
Section 10.3; 
see also Appendix \ref{a:cov}.
Here we check only invariance under $\cG_V$ (see \refeq{v-coaction}):
\beqa
c_n\to&& \tr_q(r^{-1} M^n r)=(r^{-1})^i_j (M^n)^j_k r^k_l v^l_i\nn\\
&=&S(r^i_j) (M^n)^j_k v^k_l S^2(r^l_i)=(M^n)^j_k v^k_l S(S(r^l_i)r^i_j)=
(M^n)^j_k v^k_j=c_n
\label{cn-inv}
\eeqa
as required.  As we shall see, the 
$c_n$'s for $n=1,...rank(G)-1$
fix the position of the brane configuration on the group manifold i.e. they
are quantum analogs of the $s_n$'s \refeq{sn}.

There should be another central term which is the 
quantum analog of the
ordinary determinant, which is necessary to define quantum $SU(N)$. 
It is known as
the quantum determinant, denoted by $\det_q(M)$. 
While it can be expressed as a polynomial in $c_n$'s ($n=1,..., rank(G)$),
$\det_q(M)$ is invariant under the full chiral quantum algebra.
Hence we impose the constraint
\beq
1 = \det_q(M) 
\label{qdet}
\eeq
For other groups such as $SO(N)$ and $SP(N)$, 
additional constraints (which are also invariant 
under the full chiral quantum algebra) must be imposed. 
These are known and can be 
found in the literature \cite{schupp}, but their explicit form is not
needed for the forthcoming considerations.
Appendix \ref{a:qdet} contains details about how to calculate 
$\det_q(M)$
and provides some explicit expressions.

\subsection{Realizations of RE}
\label{subsec:solutions}

In this section we find realizations (algebra homomorphisms) 
of the RE algebra \refeq{re} in terms of some other algebras.
This can be viewed as an intermediate step towards finding 
representations. We use a technique generating new solutions
out of constant solutions (i.e. trivial representations).
Thus first we consider 
 \beq
R_{21} \Mo_1 R_{12} \Mo_2 = \Mo_2 R_{21} \Mo_1 R_{12},
\label{re-o}
\eeq
where the entries of the matrices $\Mo$ are c-numbers.
Then one checks that
\beq\label{m-sol}
M=L^+M^{(0)}S(L^-)
\eeq
satisfies \refeq{re}, if the
matrices $L^\pm$ respect (see also \cite{KS}, p.285)
\beqa\label{l-alg}
&&R L^\pm_2L^\pm_1=L^\pm_1 L^\pm_2 R, \quad R L^+_2L^-_1=L^-_1L^+_2 R.
\eeqa
Notice  that $\det_q(M)=\det_q(\Mo)$ due to chiral invariance of the
q-determinant. 
Clearly the form of \refeq{m-sol} is closely related 
to the $\GcG$ invariance of the RE. Thus we have trade our original problem to
the problem of finding matrices $L^\pm$
respecting \refeq{l-alg}. Luckily this is known for a long time due 
to the famous 
work of Faddeev, Reshetikhin and Takhtajan \cite{FRT},
who noted that
\refeq{re} together with  the determinant--condition (and others 
for groups other than $SU(N)$)
provides one possible definition of
the quantized universal enveloping algebra $U_q(\mg)$.
More precisely, they showed that $L^\pm$ 
can be expressed in terms of generators of the 
$U_q(\mg)$ algebra as follows
\beqa\label{l-sol}
L^+=(id\otimes \pi)(\cR),\quad L^-=( \pi\otimes id)(\cR^{-1})
\eeqa
where $\cR=\cR_1\otimes\cR_2$ is the universal R-matrix for $U_q(\mg)$,
and $\pi$ is the defining
\rep of $U_q(\mg)$\footnote{The R-matrix of \refeq{re-i} is
  $R^i{}_j{}^k{}_l=\pi^i{}_j(\cR_1)\pi^k{}_l(\cR_2)$.}. In order to be more
transparent, we write the component form of \refeq{l-sol}:
 $(L^+)^i_j=\cR_1\pi(\cR_2)^i_j$, $(L^-)^i_j=\cR^{-1}_2\pi(\cR^{-1}_1)^i_j$.
One can also show that $SL^-$ (which we shall need later) is
\beq
SL^-=( \pi\otimes id)(\cR).
\eeq
The reason why \refeq{l-sol} respect \refeq{l-alg} is
the YBE equation for $\cR$, written in several equivalent forms
\beqa
\cR_{12}\cR_{13}\cR_{23}&=&\cR_{23}\cR_{13}\cR_{12}\\
\cR_{13}\cR_{23}\cR_{12}^{-1}&=&\cR_{12}^{-1}\cR_{23}\cR_{13}\\
\cR_{13}^{-1}\cR_{23}^{-1}\cR_{12}&=&\cR_{12}\cR_{23}^{-1}\cR_{13}^{-1}.
\eeqa
The action of $1\otimes \pi\otimes \pi$
in the first line, $\pi\otimes 1\otimes \pi$ in the
second line, and $\pi\otimes \pi\otimes 1$ in the third line 
immediately produces \refeq{l-alg}.
It is useful to realize that $L^+$ are lower triangular matrices with 
$X^+_\a$'s 
below the diagonal, and $L^-$ are upper triangular matrices with  
$X^-_\a$'s 
above the diagonal. Explicitly,  for $sl_2$ one has
\newcommand\matt[4]{\left(\barr{cc}
#1 & #2  \\#3 & #4 
\earr\right)}
\beq
L^+=\matt {q^{H/2}} 0 {q^{-\half}\la X_+} {~~~q^{-H/2}},\quad 
L^-=\matt {q^{-H/2}} {~~~ - q^{\half}\la X_-} 0 {q^{H/2}}
\eeq
The form of the solution \refeq{l-sol} shows that $M$ generates a 
sub-algebra of
$U_q(\mg)$. The sub-algebra depends on 
$\Mo$. We will not discuss the most general $\Mo$ here
(see e.g. \cite{kss}), 
but consider only the most obvious solution, which is a
diagonal matrix. The specific values  of the diagonal entries do
not change the algebra generated by the elements of $M$, because they 
simply multiply some entries of $M$. 
The other,
non-diagonal $\Mo$'s presumably also correspond to some branes: we hope to
come back to this issue in a future paper. 
Using \cite{FRT}, we conclude that the algebra
generated by the elements of $M$ is essentially
$U_q(\mg)$.
As we will see, choosing a definite \rep of $U_q(\mg)$ then
corresponds to choosing a brane
configuration, and determines the algebra of function on the brane.
To be explicit, we give the solution for $\mg = sl_2$ and $\Mo=\diag(1,1)$:
\beq
M=L^+\Mo S(L^-)=\matt {q^{H}}{q^{-\half}\la q^{H/2} X_-}
{q^{-\half}\la X_+\ q^{H/2}} {~~~~q^{-H}+q^{-1}\la^2X_+X_- }
\eeq
One can verify that $\det_q(M)=1$, according to \refeq{qdet2}.

\subsection{Covariance}
\label{subsec:cov}

We show in Appendix \ref{a:cov} that for any solutions
of the form $M=L^+ M^{(0)} S(L^-)$ where $M^{(0)}$ is a constant solution
of the RE, the ``vector'' rotations \refeq{q-adj} 
can be realized as quantum adjoint action:
\beq
 u \trr M^i_j = \pi^i_k(S u_1) M^k_l \pi^l_j(u_2)
              = u_1\; M^i_j\; S u_2\;\; \in \cM
\label{cov}
\eeq
for $u \in \cM$, where $\pi()$ is the defining \rep $V_N$ of $U_q(\mg)$.
Here we consider $\cM \subset U_q(\mg)$ so that 
$\Delta(u) = u_1 \tens u_2$ is defined in $U_q(\mg) \tens U_q(\mg)$,
nevertheless the rhs is in $\cM$. 
This is as it should be in a quantum theory: the action of a symmetry
is implemented by a conjugation in the algebra of operators.
It will be essential later to do the harmonic analysis on the branes.

\subsection{Reality structure}
\label{sec:star}

An algebra $\cM$ can be considered as a
quantized (algebra of complex-valued functions on a) space
only if it is equipped with a $*$-structure, i.e. an anti-linear 
(anti)-involution.
For classical unitary matrices, the 
condition would be $M^{\dagger} = M^{-1}$.
To find the correct quantum version is a bit tricky;
we determine it 
by requiring that on finite-dimensional representations
of $M = L^+ S L^-$ (i.e. on the branes, see below),
the $*$ will become the usual matrix adjoint.
In term of the generators of $U_q(\mg)$, this means that 
$(X_i^\pm)^* = X_i^\mp$, $H_i^* = H_i$. 
In the $SU(2)$ case, this leads to
\beq
\left(\begin{array}{cc} a^* & b^* \\ c^* & d^* \end{array}\right) = 
\left(\begin{array}{cc} a^{-1} & -q c a^{-1}\\ 
  -q a^{-1} b\;\; & q^2 d + a-q^2 a^{-1}\end{array}\right);
\label{su2-star}
\eeq
$a^{-1}$ indeed exists on the irreps of $\cM$ considered here.
A closed form for this star structure for general $\mg$ 
could also be given, but shall be omitted here.

\section{Representations of $\cM$ and quantum $D$--branes}

By construction, the $M^i_j$ can be considered as quantized coordinate
functions on G, defining some kind of quantization of the manifold $G$.
However, we are interested here in the quantization of the
orbits $\cC(t_\la)$, which are submanifolds of $G$.
We claim that they are described by irreps (fixed by the set of Casimirs)
$\pi_\la: \cM \to Mat(V_\la,\C)$ of $\cM$. Indeed, the map
$\pi_\la$ can be considered as the dual of the embedding map
$\cC(t_\la) \hookrightarrow G$. This will allow us to make statements on the
location of the branes in G.

Consider an irreducible representation of $\cM$.
The Casimirs $c_n$ \refeq{central2-n} 
then take distinct values which can be calculated.
Moreover, they are invariant under (vector) rotations
as shown in \refeq{cn-inv}.
In view of their form \refeq{central2-n}, 
this suggests that an irrep of $\cM$ should be considered as 
quantization of (the algebra of functions on) some 
conjugacy class $\cC(t_\la)$,
the position of which is determined by the values of the Casimirs $c_n$.

We will show that the irreps of $\cM$ 
describe indeed precisely the {\em stable} D-branes.
Since the algebra $\cM$ is\footnote{more precisely the semi-simple quotient 
of $\cM$, see Section \ref{subsec:harmonics}} 
the direct sum of the corresponding 
representations, 
the whole group manifold is recovered
in the limit $k \rightarrow \infty$ where the branes become dense.
To confirm this interpretation, we will calculate
the position of the branes on the group manifold, and study their
geometry by performing the
harmonic analysis on the branes, i.e. by determining the set of harmonics.

Here we shall consider only 
those representations of $M$ 
which arise from $M^{(0)} = 1$ i.e. $M = L^+ SL^-$ \refeq{m-sol}. 
Then the \reps
of the algebra $\cM$ coincide
with those of $U_q(\mg)$, which are
largely understood,
although quite complicated at roots of unity.
The fact relevant for us is that
representations $V_\la$ of $U_q(\mg)$ 
with $\la\in P_k^+$ 
have the following properties:
\begin{itemize}
\item
they are unitary, i.e. $*$ reps of $\cM$
with respect to the $*$ structure of 
Section \ref{sec:star} (see \cite{unitary})
\item their quantum-dimension $\dim_q(V_{\la})  = \tr_{V_\la}(q^{2 H_\rho})$
given in \refeq{qdim} is positive \cite{ch-p}
\item $\la$ corresponds precisely to the integrable 
modules of the affine Lie algebra $\mgh$ which governs the 
CFT.
\end{itemize}
The \reps belonging to the boundary of $P_k^+$ will correspond to the
degenerate branes.

Having characterized 
the admissible representations $V_{\la}$, we propose that
{\bf the representation of $\cM$ on $V_{\la}$ for $\la \in P_k^+$
is a quantized or ``fuzzy'' D--brane, denoted by $D_\la$}. 
It is an
algebra of maps from $V_{\la}$ to $V_{\la}$ which transforms under 
the quantum adjoint action \refeq{cov}
of $U_q(\mg)$.
For ``small'' weights\footnote{see Section \ref{subsec:harmonics}} $\la$, 
this algebra coincides with $Mat(V_{\la})$.
There are some modifications for ``large'' weights
 $\la$  because $q$ is a root of 
unity, which will be discussed in Section \ref{subsec:harmonics}. The reason
is that $Mat(V_{\la})$ then contains unphysical degrees of freedom which 
should be truncated.

A first justification is that there is indeed a one--to--one correspondence
between the (untwisted) branes in string theory and these quantum branes, 
since both are labeled by $\la \in P_k^+$.
To give a more detailed comparison, we
calculate the traces (\ref{central2-n}), derive a characteristic equation,
and then perform the harmonic analysis on $D_\la$.
Furthermore, the energy \refeq{crdymss} of the branes in string theory 
will be recovered precisely in terms of the quantum dimension.

\subsection{Value of the central terms}
\label{subsec:pos}
 
The values of the Casimirs $c_n$ on  $D_\la$ are  
calculated in Appendix \ref{a:casimirs}:
\beqa
c_0 &=& \tr_{V_N}\; (q^{-2 H_\rho}) 
        =\dim_q(V_N), \\
c_1(\la) &=& \tr_{V_N}\; (q^{2 (H_\rho +H_\la)}), \label{c1}\\
c_n(\la) &=& \sum_{\nu \in V_N;\; \la + \nu \in P_k^+} 
   q^{2n((\la + \rho)\cdot \nu - \la_N\cdot\rho)} \; 
   \frac{\dim_q(V_{\la+\nu})}{\dim_q(V_{\la})},\quad n \geq 1. \label{cn}
\eeqa
Here $\la_N$ is the highest weight of the defining \rep $V_N$, and 
the sum in \refeq{cn} goes over all $\nu \in V_N$ 
such that $\la + \nu$ lies in $P_k^+$. $c_0$ is $\la$-independent
uninteresting number.

The value of $c_1(\la)$ agrees with the corresponding
value \refeq{sn}
of $s_1$ on the classical conjugacy classes $\cC(t_{\la})$.
For $n \geq 2$, the values of $c_n(\la)$ agree only approximately
with $s_n$ on $\cC(t_{\la})$,
more precisely they agree
if $\frac{{}_q\dim(V_{\la+\nu})}{{}_q\dim(V_{\la})} \approx 1$,
which holds provided $\la$ is large 
(hence $k$ must be large too).
In particular, this holds for branes which are not ``too close'' to
the unit element.
This discrepancy for small $\la$ is perhaps not too surprising, since the
higher--order Casimirs are defined in terms of non-commutative 
coordinates and are hence subject to  operator--ordering
ambiguities.

We should emphasize here that the agreement of the 
values of $c_n$ with their
classical counterparts \refeq{sn} shows 
that the $M$'s are indeed very reasonable variables to describe the branes.

Hence we see that the positions and the ``size'' of the branes
essentially agree with the results from string theory. 
In particular, their size shrinks to zero
if $\la$ approaches a corner of $P_k^+$,
as can be seen easily in the $SU(2)$ case \cite{PS}: 
as $\la$ goes from $0$ to $k$,
the branes start at the identity $e$, grow up to the equator,
and then shrink again around $-e$.
We will see that the algebra of functions on $D_\la$ precisely reflects 
this behavior; 
however this is more subtle and will be discussed below.
All of this is fundamentally 
tied to the fact that $q$ is a root of unity.

Furthermore, the quantum dimension of the \rep space $V_{\la}$
is
\beq
\dim_q(V_\la) = \tr_{q}(1) = \tr_{V_\la}(q^{2 H_\rho}) = 
   \prod_{\a > 0} \frac{\sin(\pi\frac{\a\cdot(\la + \rho))}{k + \gcheck})}
                     {\sin(\pi\frac{\a\cdot\rho}{k + \gcheck})}.
\label{qdim}
\eeq
The last equality above follows from Weyl's character formula.
According to the interpretation \refeq{energy} it should be the energy of
the D-brane, and this is indeed the case (see \refeq{crdymss}).

Finally, we show in Appendix \ref{a:chareq} that
the generators of $\cM$ satisfy the following characteristic 
equation on $D_{\la}$:
\beq
P_\la(M) = \prod_{\nu \in V_N} 
                      (\; M -q^{2 (\la + \rho)\cdot\nu-2 \la_N\cdot\rho}) = 0.
\label{char-poly-Y}
\eeq 
Here the usual matrix multiplication of the $M^i_j$ is understood.
Again, this (almost) matches with the classical version \refeq{charpoly-class}.

\subsection{The space of harmonics on $D_\la$.}
\label{subsec:harmonics}

As discussed in Section 3, we must finally match the space of functions
or harmonics on $D_\la$ with the ones on $\cC(t_{\la})$,
up to some cutoff. Using covariance \refeq{cov}, 
this amounts to calculating the decomposition 
of $\cM$  generated by  \refeq{m-sol} characterized by $\la\in
P_k^+$ under the quantum adjoint action of $U_q(\mg)$ \refeq{q-adj}.
i.e. decomposing $V_{\la} \tens V_{\la}^*$ under $U_q(\mg)$.
To simplify the analysis, we
assume first that $\la$ is not too 
large\footnote{roughly speaking if $\la = \sum n_i \La_i$, 
then $\sum_i n_i < \frac 12 (k+\gcheck)$.}, so that 
this tensor product is completely reducible. Then
$D_\la$ coincides with the matrix algebra acting
on $V_\la$,
\beq
D_\la \cong Mat(V_\la) =  V_{\la} \tens V_{\la}^*
 \cong \oplus_{\mu} N_{\la \la^+}^{\mu} \; V_{\mu},
\label{little-rich}
\eeq
where $N_{\la \la^+}^{\mu}$ are the usual fusion rules of $\mg$
which can be calculated explicitly using formula \refeq{racah-speiser}.
Here $\la^+$ is the conjugate weight to $\la$, so that 
$V_{\la}^* \cong V_{\la^+}$.
This has a simple geometrical meaning if 
$\mu$  is small enough (smaller than all {\em nonzero}
Dynkin labels of $\la$, roughly speaking; 
see Appendix \ref{subsec:harmonic-brane} for details):
 then
\beq
N_{\la \la^+}^{\mu} =  mult_{\mu^+}^{(K_{\la})},
\label{Dla-decomp}
\eeq
where $K_{\la}\subset G$ is the stabilizer group\footnote{which acts
by the (co)adjoint action on weights} of $\la$,
and $mult_{\mu^+}^{(K_{\la})}$ is the dimension of the subspace of
$V_\mu^*$ which is invariant under $K_{\la}$.
This is proved in Appendix \ref{subsec:harmonic-brane}.
Note in particular that the mode structure (for small $\mu$)
does not depend on the particular value of $\la$, only 
on its stabilizer $K_{\la}$. 
Comparing this with the decomposition \refeq{FCt-decomp} 
of $\cF(\cC(t_\la))$, we see that indeed
\beq
D_\la \cong \cF(\cC(t_{\la}'))
\label{D-C-match}
\eeq
up to some cutoff in $\mu$, 
where $t_{\la}' = \exp(2 \pi i \frac{H_\la}{k + \gcheck})$.
This differs slightly from  \refeq{stable-branes}, by a shift
$\la \rightarrow \la + \rho$. It implies that 
degenerate branes do occur in the our quantum algebraic description, 
because $\la$ may be invariant under a nontrivial subgroup
$K_{\la} \neq T$.
These degenerate branes have smaller dimensions than 
the regular ones. An example for this is fuzzy 
$\C P^N$, which will be discussed in some detail below.

Here we differ from \cite{FFFS}
who identify only regular $D$--branes in the CFT description,
arguing that $\la + \rho$ is always regular.
This is due to a particular limiting procedure for 
$k \rightarrow \infty$ which was chosen in \cite{FFFS}. 
We assume $k$ to be large but finite, 
and find that degenerate branes do occur. This is in agreement with
the CFT description of harmonics on $D_\la$,
as will be discussed below.
Also, note that \refeq{D-C-match}
reconciles the results \refeq{c1}, \refeq{sn} on the position of the branes
with their mode structure as found in CFT.

Now we consider the general case where 
the tensor product $Mat(V_\la) = V_{\la} \tens V_{\la}^*$
may not be completely reducible. Then
$Mat(V_\la) = V_{\la} \tens V_{\la}^*$
contains non-classical \reps with vanishing quantum dimension,
which have no obvious interpretation.
However, there is a well--known remedy: one can replace the 
full tensor product by the so-called ``truncated tensor product'' 
\cite{b-fuchs},
which amounts to discarding
\footnote{note that the calculation
of the Casimirs in Section \ref{subsec:pos} is still valid, because
$V_\la$ is always an irrep} the \reps with $dim_q = 0$.
This gives a decomposition into irreps
\beq
D_\la  \cong V_{\la} \obar{\tens} V_{\la}^*
 \cong \oplus_{\mu \in P_k^+} \obar{N}_{\la \la^+}^{\mu} \; V_{\mu}
\label{little-rich-trunc}
\eeq
involving only modules $V_{\mu}$ of positive quantum dimension. 
These 
$\obar{N}_{\la \la^+}^{\mu}$
are known to coincide with the fusion rules for integrable modules
of the affine Lie algebra $\mgh$ at level $k$, and
can be calculated explicitly.
These fusion rules in turn coincide (see e.g. \cite{FFFS}) with
the multiplicities of harmonics on the D-branes
in the CFT description, 
i.e. primary (boundary) fields. 

We conclude that the structure of
harmonics on $D_\la$, \refeq{little-rich-trunc}
is in complete agreement with the CFT results. 
Moreover, it is known (see also \cite{FFFS})
that the  structure constants of the corresponding boundary operators
are essentially given by the $6j$ symbols of $U_q(\mg)$, which in turn are 
precisely the structure constants of the algebra of functions 
on $D_\la$, as explained in \cite{fuzzyqsphere}.
Therefore our quantum algebraic description not only 
reproduces the correct set of boundary fields, but also 
essentially captures their algebra in (B)CFT.

Finally, it is interesting to note 
that branes $D_\la$ which are  ``almost'' degenerate
(i.e. for $\la$  near some boundary of $P_k^+$) have only few
modes $\mu$  in some directions\footnote{this is just the condition on
$\mu$ discussed before \refeq{Dla-decomp}}
and should therefore be interpreted
as degenerated branes with ``thin'', but finite walls. 
They interpolate between branes of different dimensions.

\section{Examples}
\label{sec:examples}
\subsection{Fuzzy $\C P^{N-1}_q$}

Particularly interesting examples of degenerate conjugacy classes are  
the complex projective spaces $\C P^{N-1}$.
We shall demonstrate the scope of our general results by extracting some 
explicit formulae for this special case.
This gives a $q$-deformation of the fuzzy $\C P^{N-1}$ discussed 
in \cite{bala,stroh}.

We first give a more explicit description of branes on $SU(N)$. 
Let $\la^a = (\la^a)^\a{}_{\dot{\b}}$ for $a = 1, 2, .., N^2-1$
be the $q$-deformed Gell-Mann
matrices, i.e. the intertwiners
$(N) \tens (\obar{N}) \to (N^2-1)$ for $U_q(su(N))$. 
We can then parameterize the matrix $M$ ($= L^+ SL^-$ acting on 
$V_\la$) as
\beq
M = \sum_a \xi_a \la^a + \xi_0 \la^0
\label{M-param}
\eeq
where we set $\la^0 \equiv {\bf 1}$. The $\xi_a$ will be generators 
of a non-commutative algebra. 
The matrices $\la_a$ satisfy
\beq
\la^a \la^b = \frac1{\dim_q(V_N)} g^{ab} + 
                 (d^{ab}{}_c  +  f^{ab}{}_c) \la^
\label{gell-mann-id}
\eeq
where $g^{ab}$, $d^{ab}{}_c$ and  $f^{ab}{}_c$ 
are invariant tensors in a suitable
normalization, and $\tr_q(\la^a) = 0$ (for $a\neq 0$).
We can now express the Casimirs $c_n$ \refeq{cn} in terms of the
new generators:
\beqa
c_1 &=& \tr_q(M) = \xi_0\; \dim_q(V_N), \label{brane-pos}\\
c_2 &=& g^{ab}\; \xi_a \xi_b + \xi_0^2\; \dim_q(V_N), \label{brane-radius}\\
\eeqa
etc, which are numbers on each $D_\la$. 
An immediate consequence of \refeq{brane-pos} is
\beq
[\xi_0, \xi_a] = 0
\label{brane-CR-1}
\eeq
for all $a$. One can show furthermore
that the reflection equation \refeq{re}, which is equivalent 
to the statement that the ($q$-)antisymmetric part of $MM$ vanishes, 
implies that
\beq
f^{ab}{}_{c}\; \xi_a \xi_b = \a \; \xi_0 \xi_c.
\label{brane-CR}
\eeq
On a given brane $D_\la$, $\xi_0$ is a number determined by \refeq{brane-pos}, 
while $\a$ is a (universal) constant 
which can be determined explicitly, as indicated below.

\refeq{brane-CR-1} and \refeq{brane-CR} hold for all branes $D_\la$.
Now consider $\C P^{N-1} \cong SU(N)/ U(N-1)$, which
is the conjugacy class through $\la = n \La_1$
(or equivalently $\la = n \La_N$) where $\La_i$ 
are the fundamental weights; 
indeed, the stabilizer group for $n\La_1$ is $U(N-1)$.
The quantization of $\C P^{N-1}$ is therefore the brane $D_{\la}$.
It is characterized by a further relation among the generators $\xi_a$,
which has the form
\beq
d^{ab}{}_{c}\; \xi_a \xi_b = \b_n\; \xi_c
\label{CPN-rel}
\eeq
where the number $\b_n$  can be determined explicitly as indicated below.
For $q=1$, these relations reduce to the ones given in \cite{bala}.
\refeq{CPN-rel} can be derived using the results in 
Section \ref{subsec:harmonics}:
It is easy to see using \refeq{racah-speiser} that 
\beq
D_{n \La_1} \cong \oplus_n (n,0,..., 0,n)
\eeq
up to some cutoff,
where $(k_1, ..., k_N)$ denotes the highest-weight \rep with Dynkin labels
$k_1, ..., k_N$. 
In particular, all multiplicities are one. This implies that 
the function $d^{ab}{}_{c} \xi_a \xi_b$ on $D_{n \La_1}$ must 
be proportional to $\xi_c$, because it transforms as $(1,0,..., 0,1)$
(which is the adjoint). Hence \refeq{CPN-rel} follows.

The constant $\a$ in \refeq{brane-CR} can be calculated
either by working out RE explicitly, or 
by specializing \refeq{brane-CR} for $D_{\La_1}$. 
We shall only indicate this here:
On $D_{\La_1}$, $\xi_a = c \la_a$ for some $c \in \C$. 
Plugging this into \refeq{brane-CR}, one finds
$c\, f^{ab}{}_{c}\; \la_a \la_b = \a \;\xi_0 \la_c$,
and  $c^2\; g^{ab}\; \la_a \la_b + \xi_0^2\; \dim_q(V_N) = c_2$. 
Calculating $\xi_0$ and the  Casimirs  explicitly on $D_{\La_1}$,
one obtains $\a$ which vanishes as $q\rightarrow 1$. 
Similarly using the explicit value of $c_3$ given in Section \ref{subsec:pos},
one can also determine $\b_n$.
Alternatively, they be calculated using creation - and
annihilation operator techniques of \cite{grosse1}, \cite{fuzzyqsphere}.

In any case, we recover the relations of fuzzy $\C P^{N-1}$
as given in \cite{bala} in the limit $q \to 1$. As an algebra, 
it is in fact identical to it,
as long as $k$ is sufficiently large.

\subsection{$G=SU(2)$ model}

In this section we shall show how one can recover the results 
of \cite{PS} from the 
general formalism we discussed so far.
The \rep of the RE given by $L^\pm$ operators and $\Mo=\diag(1,1)$ is
\beq
M=L^+\Mo S(L^-)=\left(\barr{cc}{q^{H}}&{q^{-\half}\la q^{H/2} X_-}\\
{q^{-\half}\la X_+\ q^{H/2}}& {~~~~q^{-H}+q^{-1}\la^2X_+X_- }\earr\right)
\eeq
Let us parameterize the $M$ matrix as
\beq\label{m-para}
M =\left(\barr{cc} 
M^4- i M^0& -i q^{-3/2}\sqrt{[2]}M^+\\ iq^{-1/2}\sqrt{[2]}M^-& M^4+iq^{-2}M^0 \earr\right)
\eeq
(cp. \refeq{M-param}), then RE is equivalent to 
\beqa\label{eq-m}
[M^4, M^l] = 0,\quad
\ep_{ij}^l \; M^i M^j = i (q-q^{-1}) M^4 M^l 
\eeqa
In order to calculate the central terms 
we need 
\beq
v=\pi(q^{-2H_\rho}) = \pi(q^{-H}) =\diag(q^{-1},q)
\eeq
so that (using \refeq{central2-n},\refeq{qdet2})
\beqa
c_1&=&\tr_q(M)=q^{-1} a+ q d=[2]\,M^4\\
c_2 &=&\tr_q(M^2)= [2]\;((M^4)^2 - q^{-2} g_{ij} M^i M^j)  \\
\det_q(M)&=&(M^4)^2+(M^0)^2-q^{-1}M^+M^--q\,M^-M^+
 =  (M^4)^2 +g_{ij} M^i M^j.
\eeqa
Only $\det_q(M)$ is invariant under $\gcg$.
The explicit value of $M^4=c_1/[2]$ is obtained from
\beq
M^4=\frac1{[2]}(q^{-1}a+q\,d)=\frac1{[2]}(q^{H-1}+q^{-(H-1)}+ \la^2 X_+X_-)
\eeq
which is proportional to the 
standard Casimir of $U_q(su(2))$. 
On the n-th brane $D_n$,  $H$ takes the value $-n$ on the
lowest weight vector, thus
$M^4=\cos(\frac{(n+1)\pi}{k+2})/\cos(\frac{\pi}{k+2})$.
If the square of radius of the quantum $S^3$ is chosen to be $\det_q(M)=k$ 
(which is the value given by the supergravity solution for the background),
$g_{ij} M^i M^j$
leads to the correct formulae for the square of the radius of the n-th branes.

\section{Conclusion}

In this paper we propose a simple and compact description 
of all (untwisted) D-branes on group manifolds $G$ based on
the reflection equation RE. 
The model 
can be viewed as a finite matrix model in the spirit of the
non-abelian DBI model of D0-branes \cite{myers},
but contrary to the latter it yields results well beyond 
the $1/k$ approximation.
In fact,  the model properly describes all branes on the group manifold
regardless of their positions. 
This covers an astonishing wealth of data on the configurations
and properties of  branes such as their positions and spaces 
of functions, which are shown to be in very good 
agreement with the CFT data. 
It also shows that $M$ is a very reasonable
variable to describe the branes. 
Our construction also sheds light on the fact that the energies
of these branes are given by so-called quantum dimensions. 

The branes are uniquely given by
certain ``canonical'' irreps of the RE
algebra, and their world-volume can be interpreted as quantum manifolds. 
The characteristic feature of our construction is
the covariance of RE under
a quantum analog of the group of isometries $G_L \times G_R$ of $G$.
A given brane configuration breaks it 
to the diagonal (quantum) $\cG_V$, an analog 
of the classical vector symmetry $G_V$.

Let us also mention that the methods worked out in this paper
might also serve as tools describing branes in RR background. E.g. it is known
\cite{jsr} that 
for $G=SU(2)=S^3$ the
dynamics of branes is very similar for both NSNS and RR backgrounds. 

It should be clear to the reader that 
the present paper does not cover all aspects of branes physics
on group manifolds.
For example, we
did not study all \reps of RE, only the most obvious ones
which are induced by the algebra map $RE \to U_q(\mg)$. 
There exist other \reps of RE, some of which can be investigated 
using the  technique in Section \ref{subsec:solutions},
some of which may be entirely different. One may hope that all
of 
the known $D$-branes on groups, including those not discussed here such as 
twisted branes
or ``type B''-branes, can be described in this way.
We plan to investigate this further in a future publication.
Moreover, we did not touch here the dynamical aspects of $D$-branes, 
such as their excitations and interactions. For this it may be necessary to
extend the algebraic content  
presented here, and the well-developed theory of quantum groups
may become very useful.

The paper has also an interesting mathematical side.
The general construction of quantized branes
presented here yields immediately a variety of specify examples of 
finite (``fuzzy'') quantum spaces, including $\C P^N_q$. They may serve
as useful testing grounds for noncommutative field theories,
which can be defined in a very clean way on fuzzy spaces, being finite.
This
should lead to further insights into the problems encountered
recently on other spaces with star-products.
Some work in that direction can be found for example in 
\cite{grosse1,fuzzyloop,vaidya}.

\section*{Acknowledgements}

We would like to thank 
H.Grosse, B. Jurco, J.Madore, P. Schupp, A. Strohmaier and J. Wess 
for useful discussions. We would also like to thank the 
Laboratoire de Physique Th\'eorique, Universit\'e de Paris-Sud at Orsay, the
Ludwig--Maximilians--Universit\"at M\"unchen and MPI f\"ur Gravitationsphysik,
Albert-Einstein-Institut at Potsdam for hospitality.  
This work was supported in part by the 
Polish State Committee for Scientific Research (KBN) under contract
5 P03B 150 20 (2001-2002), and in part by a DFG fellowship.

\begin{appendix}

\section{Appendix: technical details}

\subsection{Some properties of $U_q(\mg)$}
\label{a:basics}

We collect here some definitions, in order to establish the notations.
We basically follow the conventions of \cite{majid}.
$\mg$ is a simple Lie algebra,
with Cartan matrix $A_{ij}=2\frac{\a_i\cdot\a_j}{\a_j\cdot\a_j}$.
The generators $X^{\pm}_i, H_i$ of $U_q(\mg)$ satisfy the relations
\beqa
\;[H_i, H_j]      &=& 0,  \quad
\;[H_i, X^{\pm}_j] = \pm A_{ji} X^{\pm}_j,     \\
\;[X^+_i, X^-_j]   &=& \d_{i,j}
                      \frac{q^{d_i H_i} -q^{-d_i H_i}}{q^{d_i}-q^{-d_i}}
                    = \d_{i,j} [H_i]_{q_i}
\label{UEA}
\eeqa
where $q_i = q^{d_i}$.
Comultiplication and antipode are defined by
\beqa
\Delta(H_i)          &=& H_i \tens 1 + 1 \tens H_i,  \qquad
\Delta(X^{\pm}_i) =  X^{\pm}_i \tens q^{d_i H_i/2} + q^{-d_i H_i/2}
\tens  X^{\pm}_i, \nn\\
S(H_i)    &=& -H_i, \qquad
S(X^{\pm}_i)  = -q^{\pm d_i} X^{\pm}_i.
\label{coproduct-X}
\eeqa
The coproduct is conveniently written in Sweedler-notation as
$\Delta(u) = u_1 \tens u_2$, for $u\in U_q(\mg)$,
where a summation is implied.
It is easy to verify that
$S^2(u) = q^{2 H_\rho} u q^{-2 H_\rho}$ for all $u \in U_q(\mg)$,
where $\rho = \frac 12 \sum_{\a>0} \a$ is the Weyl vector.
This is used in the definition of the quantum traces \refeq{central2-n},
\refeq{qdim}.

\subsection{The dual symmetries of the reflection equation.}
\label{a:symmetries}

Let $\cG$ and $\cU$ be Hopf algebras. An algebra $\cM$ is a (left)
$\cU$-module algebra if there is an action 
$\trr: \cU \times \cM \to \cM$
such that $u \trr (m n) = (u_1 \trr m) (u_2 \trr n)$ 
and $(uv)\trr m = u \trr(v\trr m)$
for $m,n \in \cM$ and $u,v \in \cU$,
where $\Delta(u) = u_1 \tens u_2$.  
$\cM$ is a  (right) $\cG$-comodule algebra
if there is a coaction $\nabla: \cM \to \cM \tens \cG$ which is 
an algebra map and satisfies
$(id \tens \Delta)\nabla(m) = (\nabla \tens id)\nabla (m)$.
These are dual concepts: if $\cG$ and $\cU$ are dually paired Hopf algebras
(see e.g. \cite{majid}), then a (right) $\cG$ - comodule algebra $\cM$
with coaction $\nabla$
is automatically a (left) $\cU$ -- module algebra
by 
\beq
u \trr M = \langle\nabla(M),u\rangle
\eeq
where $\langle m\tens a,u\rangle = m\langle a,u\rangle$,
and vice versa.

This is exactly our situation: The Hopf algebra
$\cG_L \tens^\cR \cG_R$ \refeq{gg-algebra} is dual to 
$U_q(\mg_L \times \mg_R)_\cR$, which as an 
algebra is the usual tensor product
$U_q(\mg_L) \tens U_q(\mg_R)$, but has the twisted Hopf structure
\beqa
\Delta_{\cR}: U_q^L \otimes U_q^R &\rightarrow&
(U_q^L \otimes U_q^R) \otimes 
     (U_q^L \otimes U_q^R), \non
   u^L \otimes u^R &\mapsto& {\cF}(u^L_1 \otimes u^R_1) \otimes (u^L_2 \otimes u^R_2) {\cF}^{-1}
\label{coprod-so(4)-new}
\eeqa
with ${\cF} = 1\tens \cR{}^{-1}\tens 1$.
This is a special case of a ``Drinfeld twist'',
which provides also a corresponding antipode and counit (and $\cR$ - matrix;
for the general theory of twisting    
we refer to \cite{ch-p}, and \cite{majid}, Section 2.3).
The dual evaluation $\langle, \rangle$ between $\cG_L \tens^\cR \cG_R$
and $U_q(\mg_L \times \mg_R)_\cR$
is defined componentwise,
using the standard dualities of $\cG_{L,R}$ with $U_q(\mg_{L,R})$.

The action of $U_q(\mg_L \times \mg_R)_\cR$ on $\cM$ which is dual to 
\refeq{st-coaction} then comes out as
\beq
(u_l\tens u_R) \trr M^i_j = M^k_l \langle S(s^i_k) t^l_j,u_l\tens
u_R\rangle
= \pi(S u_L) M \pi(u_R),
\eeq
as in \refeq{UU-action}.
Moreover, there is a Hopf-algebra map 
$u \in U_q(\mg_V) \to \Delta(u) \in U_q(\mg_L \times \mg_R)_\cR$
where $\Delta$ is the usual coproduct. This
defines the vector sub-algebra $U_q(\mg_V)$. 
It induces on $\cM$ the action \refeq{q-adj}, which is 
again dual to the coaction \refeq{v-coaction}.

At roots of unity, these dualities are somewhat more subtle. We will 
not worry about this, because covariance of
the reflection equation under $U_q(\mg_L \times \mg_R)_\cR$ can also be 
verified directly.

\subsection{The harmonics on the branes}
\label{subsec:harmonic-brane}

\paragraph{The modes on $\cC(t)$} \refeq{FCt-decomp}

Consider the map
\beqa
G/K_t &&\rightarrow \cC(t), \nn\\
g K_t  &&\mapsto  g t g^{-1}  \nn
\eeqa
which is clearly well--defined and bijective.
It is also compatible with the group actions, in the sense that 
the adjoint action of $G$ on $\cC(t)$ translates into the {\em left} action
on $G/K_t$. Hence we want to decompose functions on 
$G/K_t$ under the left action of $G$. 

Functions on 
$G/K_t$ can be considered as functions on $G$ which are invariant under the 
{\em right} action of $K_t$, and this correspondence is one-to-one
(because this action is free). Now
the Peter-Weyl theorem states that the 
space of functions on $G$ is isomorphic as a bimodule to 
\beq
\cF(G) \cong  \bigoplus_{\la \in P^+} \; V_{\la} \tens V_{\la}^*.
\label{FG-decomp}
\eeq
Here $\la$ runs over all dominant integral weights, and
$V_{\la}$ is the corresponding highest-weight module. 
Let $mult_{\la^+}^{(K_t)}$ be the dimension of the 
subspace of $V_{\la}^* \equiv V_{\la^+}$ which is
invariant under the action of $K_t$. Then
\beq
\cF(\cC(t)) \cong  \bigoplus_{\la \in P^+} \; mult_{\la^+}^{(K_t)}\; V_{\la}
\eeq
follows.

\paragraph{The modes on $D_{\la}$ and proof of}
(\ref{Dla-decomp})

We are looking for the Littlewood--Richardson 
coefficients $N_{\la \la^+}^{\mu}$ in the decomposition
\beq
V_{\la} \tens V_{\la}^* \cong \oplus_{\mu} N_{\la \la^+}^{\mu} \; V_{\mu}
\eeq
of $\mg$ - modules.
Now we use $N_{\la \la^+}^{\mu} = N_{\la \mu^+}^{\la}$
(because $N_{\la \la^+}^{\mu}$ is given by
the multiplicity of the trivial component in 
$V_{\la} \tens V_{\la^+} \tens V_{\mu^+}$,
and so is $N_{\la \mu^+}^{\la}$).
But $N_{\la \mu^+}^{\la}$ can be calculated using the formula \cite{b-fuchs}
\beq
N_{\la \mu^+}^{\la} = \sum_{\sigma  \in W} \; (-1)^{\sigma} \;
        mult_{\mu^+}(\sigma \star \la - \la),
\label{racah-speiser}
\eeq
where $W$ is the Weyl group of $\mg$.
Here $mult_{\mu^+}(\nu)$ is the multiplicity of the weight space $\nu$
in $V_{\mu^+}$, and $\sigma \star \la = \sigma(\la + \rho) - \rho$ 
denotes the action of $\sigma$ 
with reflection center $-\rho$.
Now one can see already that for large, generic $\la$ (so that 
$\sigma \star \la - \la$ is not a weight of $V_{\mu^+}$ unless $\sigma = 1$), 
it follows that $N_{\la \mu^+}^{\la} = mult_{\mu^+}(0) = mult_{\mu^+}^{(T)}$,
which proves \refeq{Dla-decomp} for the generic case. 
To cover all possible $\la$, we proceed as follows:

Let $\mk$ be the Lie algebra of $K_{\la}$, and
$W_{\mk}$ its Weyl group; it is the subgroup of $W$ which 
leaves $\la$ invariant, generated by 
those reflections which preserve $\la$
(the $u(1)$ factors in  $\mk$ do not contribute to $W_{\mk}$).
If $\mu$ is ``small enough'', then 
the sum in \refeq{racah-speiser}
can be restricted to $\sigma \in W_{\mk}$, because otherwise 
$\sigma \star \la - \la$ is too large to be in $V_{\mu^+}$;
this defines the cutoff
in $\mu$. It holds for any given $\mu$ if $\la$ has the form
$\la = n \la_0$ for large $n \in \N$ 
and fixed $\la_0$\footnote{This constitutes 
our definition of ``classical limit''.
For weights $\la$ which do not satisfy this requirement, 
the corresponding $D$--brane $D_{\la}$ cannot be interpreted as 
``almost--classical''. Here we differ from the approach in \cite{FFFS},
which do not allow degenerate $\la_0$.}.
We will show below that 
\beq
mult_{\mu^+}^{(K_{\la})} = \sum_{\sigma  \in W_{\mk}} \; (-1)^{\sigma} \;
        mult_{\mu^+}(\sigma \star \la - \la)
\label{mult-eq}
\eeq
for all  $\mu$, which implies \refeq{Dla-decomp}.
Recall that the lhs is the dimension of the subspace of
$V_{\mu^+}$ which is invariant under $K_{\la}$. 

To prove \refeq{mult-eq}, first observe the following fact: Let
$V_{\la}$ be the highest weight irrep of some simple Lie algebra $\mk$
with highest weight $\la$. Then 
\beq
\sum_{\sigma \in W_{\mk}}\; (-1)^{\sigma}\;mult_{V_{\la}}(\sigma \star 0)
  = \d_{\la,0}
\label{trivial-rep}
\eeq
i.e. the sum vanishes unless $V_{\la}$ is the trivial 
representation; here $\mk = u(1)$ is allowed as well.
This follows again from \refeq{racah-speiser},
considering the decomposition of 
$V_\la \tens V_0$.
More generally, assume that  $\mk = \oplus_i \mk_i$ is a direct sum
of simple Lie algebras $\mk_i$, with corresponding 
Weyl group $W_{\mk} = \prod_i W_i$. 
Its irreps have the form $V = \tens_i V_{\la_i}$, where $V_{\la_i}$ 
denotes the highest weight module of $\mk_i$ with highest weight $\la_i$.
We claim that the relation
\beq
\sum_{\sigma \in W_{\mk}}\; (-1)^{\sigma}\;mult_{V}(\sigma \star 0)
  = \prod_i\; \d_{\la_i,0}
\label{trivial-mult}
\eeq
still holds. Indeed, assume that some $\la_i \neq 0$; then 
$$
\sum_{\sigma \in W_{\mk}}\; (-1)^{\sigma}\;mult_{V}(\sigma \star 0) = 
\Big(\sum_{\sigma' \in\; \prod' W_i} (-1)^{\sigma'}\Big)
\Big(\sum_{\sigma \in W_i} (-1)^{\sigma_i}
 mult_{V}(\sigma_i\star (\sigma' \star 0))\Big) = 0
$$
in self--explanatory notation. The last bracket vanishes by 
\refeq{trivial-rep}, since $(\sigma' \star 0)$ has weight $0$ with 
respect to $\mk_i$, while $V$ contains no trivial component of $\mk_i$
(notice that $\rho = \sum \rho_i$, and the operation $\star$ 
is defined component-wise).
Therefore for any (finite, but not necessarily irreducible) $\mk$--module 
$V$, the number of trivial components in $V$ is given by
$\sum_{\sigma \in W_{\mk}}\; (-1)^{\sigma}\;mult_{V}(\sigma \star 0)$.

We now apply this to \refeq{mult-eq}. Since the sum is over 
$\sigma \in W_{\mk}$, we have $\sigma(\la) = \la$ 
by definition, and $\sigma \star \la - \la  = \sigma \star 0$. Hence
the rhs can be replaced by 
$\sum_{\sigma \in W_{\mk}}\; (-1)^{\sigma}\;mult_{\mu^+}(\sigma \star 0)$.
But this is precisely the number of  vectors in $V_{\mu^+}$
which are invariant under $K_{\la}$, as we just proved.  
Notice that we use here the fact that $\mk$ contains the 
Cartan sub-algebra of $\mg$, so that the space of weights of $\mk$
is the same as the space of weights of $\mg$; therefore the 
multiplicities in \refeq{mult-eq} and \refeq{trivial-mult}
are defined consistently. This is why we had to include the case 
$\mk_i = u(1)$ in the above discussion.

To calculate the decomposition \refeq{little-rich-trunc} for all 
allowed $\la$ (with $\dim_q(V_{\la}) > 0$), 
the ordinary multiplicities in \refeq{little-rich} should be replaced with 
with their truncated versions $\obar{N}_{\la \mu^+}^{\la}$
\refeq{little-rich-trunc} corresponding
to $U_q(\mg)$ at roots of unity. 
There exist generalizations of the formula \refeq{racah-speiser}
which allow to calculate $\obar{N}_{\la \mu^+}^{\la}$
 efficiently; we refer here to the 
literature, e.g. \cite{FFFS}.

\subsection{The quantum determinant}
\label{a:qdet}
\def\tR{{\tilde R}}

Here we present a formula for the quantum determinant,
following \cite{schupp}.
First we have to introduce q-deformed 
totally (q)--antisymmetric tensors $\eps_q^{i_1 ... i_N}$ of $U_q(sl(N))$.
\beq
\eps_q^{\si(1) ... \si(N)}=(-q)^{-l(\si)} = \eps^q_{\si(1)
  ... \si(N)} 
\eeq
where $l(\si)$ is the length of the permutation $\si$. The important formula
respected by $\eps_q$ is (in notation of \refeq{re})
\beq
\eps_q^{1...N}R'_{i}=-\frac1{q}\eps_q^{1...N}
\eeq
where $R'_{i}=\hat R_{(i+1)\ i}$ and ${\hat R}^i_j{}^k_l=R^k_j{}^i_l$. With
this notation we define
\beq
\eps_q^{1...N} \det_q (M)={\cal N} (M_1R'_1R'_2...R'_{N-1})^N \eps_q^{1...N}
\eeq
where ${\cal N}$ is an arbitrary normalization constant. 
One can show that this is invariant under the chiral symmetries
$\GcG$ as in \refeq{st-coaction},
or equivalently under the  action of $\gcg$.

For $N=2$ we have
$(M_1R'_1) (M_1R'_1) \eps_q^{12}=-\frac1q M_1R'_1M_1 \eps_q^{12}$.
After using the RE relations, 
this becomes  proportional to $\eps_q^{12}$ times 
$q^{-1}(M^1_1M^2_2-q^2M^2_1M^1_2)$ thus we choose
the quantum determinant 
\beq\label{qdet2}
\det_q(M)=(M^1_1M^2_2-q^2M^2_1M^1_2)
\eeq

For other groups such as $SO(N)$ and $SP(N)$, the explicit form for 
$\eps_q^{i_1 ... i_N}$ is different, 
and additional constraints (which are also invariant 
under the chiral symmetries) must be imposed. These are known and can be 
found in the literature \cite{schupp}.

\subsection{Covariance of $M$ and central elements}
\label{a:cov}

For any numerical matrix $M^{(0)}$ (in the defining \rep of $U_q(\mg)$), 
consider
\beq
M = L^+ M^{(0)} SL^- = (\pi \tens 1)(\cR_{21})\; M^{(0)}\; 
   (\pi \tens 1)\cR_{12}.
\label{MC-explicit}
\eeq
Let $\cM \subset U_q(\mg)$  be the sub-algebra generated by the 
entries of this matrix. 
First, we note that $\cM$ is a (left) coideal sub-algebra, 
which means that $\Delta(\cM) \in U_q(\mg) \tens \cM$. This is 
verified simply by calculating the coproduct of $M$, 
\beq\label{coid}
\Delta({M}^i_l) = {L^+}^i_s {SL^-}^t_l \tens (M)^s_t.
\eeq
In particular if $M^{(0)}$ is a constant 
solution of the  reflection equation \refeq{re}, it follows by taking the 
defining \rep of \refeq{MC-explicit} that $[\pi({M}^i_j), M^{(0)}] = 0$,
and therefore $[\pi(\cM), M^{(0)}] = 0$. Then  
for any $u \in \cM \subset U_q(\mg)$,
\beqa
((\pi\tens 1)\Delta(u)) M &=& (\pi\tens 1)(\Delta(u)\cR_{21})\; M^{(0)}\; SL^- \nn\\
  &=&(\pi\tens 1)(\cR_{21}\Delta'(u))\; M^{(0)}\; SL^- \nn\\
  &=& L^+\; M^{(0)}\; (\pi\tens 1)(\Delta'(u)\cR_{12})\nn\\
  &=& L^+\; M^{(0)}\; SL^- (\pi\tens 1)\Delta(u) = M\; (\pi\tens 1)\Delta(u).
\eeqa
In the second line we used 
$\Delta'(u)\equiv u_2\otimes u_1=R \Delta(u) R^{-1}$,
in the third line,  the coideal property \refeq{coid}. 
Using Hopf algebra identities
(i.e. multiplying from left with $(\pi(S u_0) \tens 1)$ and from the right 
with $(1 \tens Su_3)$), this is equivalent to
$(1\tens u_1) M (1\tens Su_2) = (\pi(Su_1) \tens 1) M (\pi(u_2)\tens 1)$,
or 
\beq
u_1 M S u_2 = \pi(Su_1) M \pi(u_2)
\eeq
for any $u \in \cM$, as desired.
This  implies immediately that 
\beq
 u_1 \tr_q(M^n) Su_2  = \tr_q(\pi(Su_1) M^n \pi(u_2)) = \eps(u)\; \tr_q(M^n),
\eeq
or equivalently 
\beq
[u, \tr_q(M^n)] = 0
\eeq
for any $u \in \cM$. 
This proves in particular that the Casimirs $c_n$ \refeq{central2-n} 
are indeed central.

\subsection{Evaluation of Casimirs}
\label{a:casimirs}

\paragraph{Evaluation of $c_1$}

Consider the fuzzy $D$--brane $D_\la$. Then $c_1$  acts on 
the highest--weight module $V_{\la}$, and has the form
\beq
c_1 =  \tr_q(L^+ SL^-) = (\tr_q \pi \tens 1) (\cR_{21} \cR_{12}).
\eeq
Because it is a Casimir, it is enough to evaluate it on the 
lowest--weight state $|\la_-\rangle$ of $V_\la$, given by 
$\la_- = \sigma_m(\la)$ where 
$\sigma_m$ denotes the longest element of the Weyl group. 
Now the universal $\cR$ has the form
\beq
\cR = q^{H_i (B^{-1})_{ij} \tens H_j} \; (1 \tens 1 + \sum U^+ \tens U^-).
\eeq
Here $B$ is the (symmetric) 
matrix $d_j^{-1} A_{ij}$ where $A$ is the Cartan Matrix, 
$d_i$ are the lengths of the simple roots ($d_i = 1$ for $\mg = su(N)$) and
$U^+, U^-$ stands for terms in the Borel sub-algebras of
rising respectively lowering operators. Hence 
only the diagonal elements of $(SL^-)^i_j$ are non--vanishing on a
lowest--weight state, and due to the trace only the diagonal elements 
of $(L^+)^i_j$ enter. We can therefore write
\beq
c_1\; |\la_-\rangle = 
      (\tr_q \pi \tens 1) 
         (q^{2\; H_i (B^{-1})_{ij} \tens H_j}) |\la_-\rangle 
    = (\tr \;\pi \tens 1) (q^{-2 H_\rho} \tens 1)
              (q^{2\; H_i (B^{-1})_{ij} \tens H_j}) \; |\la_-\rangle
\eeq
Here 
$H_\la |\mu\rangle =(\la\cdot\mu)\;|\mu\rangle$
for any weight $\mu$. Therefore the eigenvalue of $c_1$ is
\beq
c_1 = \sum_{\mu \in V_N}\; q^{-2 \mu\cdot\rho + 2 \mu\cdot\la_-}
    = \sum_{\mu \in V_N}\; q^{2 \mu \cdot (-\rho +\la_-)}.
\eeq
Using $\sigma_m(\rho) = -\rho$, this becomes
\beq
c_1 = \sum_{\mu \in V_N}\; q^{2 (\sigma_m(\mu))\cdot(\rho +\la)}
= \tr_{V_N}\; (q^{2 (\rho +\la)})
\eeq
because the weights of $V_N$ are invariant under the Weyl group.

\paragraph{Evaluation of $c_n$ in general}

Since $c_n$ is proportional to the identity matrix on irreps, 
it is enough to calculate
$\tr_q(c_n) = \tr_{V_\la}(c_n\; q^{-2 H_\rho})$ on $V_{\la}$, 
noting that $\tr_q(1) = \dim_q(V_{\la})$
is known explicitly:
\beq
\tr_q(c_n) = (\tr_q \tens \tr_q)((\cR_{21} \cR_{12})^n)
\eeq
where the traces are over $Mat(N)$ and $Mat(V_{\la})$.
Now we use the fact that $\cR_{21} \cR_{12}$ commutes with $\Delta(U_q(\mg))$,
i.e. it is constant on the irreps of $V_{N} \tens V_{\la}$,
and observe that $\Delta(q^{-2 H_\rho}) = q^{-2 H_\rho} \tens q^{-2 H_\rho}$,
which means that the quantum trace factorizes. Hence we can
decompose the tensor product $V_{N} \tens V_{\la}$
into irreps:
\beq
V_{N} \tens V_{\la} = \oplus_{\mu\in P_k^+} V_{\mu}
\label{laLa-decomp}
\eeq
where the sum goes over all $\mu$  which have the form $\mu = \la + \nu$
for $\nu$ a weight of $V_{N}$. The multiplicities are equal
one because $V_N$ is the defining representation.
The eigenvalues of $\cR_{21} \cR_{12}$ on $V_{\mu}$ are 
known \cite{resh-1} to be
$q^{c_{\mu} - c_{\la} - c_{\la_N}}$, where $\la_N$
denotes the highest weight of $V_N$ and  $c_\la = \la\cdot(\la+2\rho)$. 
Now for $\mu = \la + \nu$, 
\beq
c_{\mu} - c_{\la} - c_{\la_N} = 2 (\la + \rho)\cdot \nu - 2\la_N\cdot\rho,
\eeq
hence the set of eigenvalues of $\cR_{21} \cR_{12}$ is
\beq
\{q^{2 (\la + \rho)\cdot\nu - 2\la_N\cdot \rho};\;\; \nu \in V_{N}\}.
\eeq
Putting this together, we obtain
\beq
\tr_q(c_n) = c_n\; \tr_{V_{\la}}(q^{-2 H_\rho}) = \sum_{\mu}\;
       q^{2n((\la + \rho)\cdot\nu - \la_N\cdot \rho)}\; 
            \tr_{V_{\mu}}(q^{-2 H_\rho})
\eeq
where the sum is as explained above.
Then \refeq{cn} follows, since 
$\tr_{V_{\mu}}(q^{-2 H_\rho}) = \dim_q(V_{\mu})$.

\subsection{Characteristic equation for $M$.}
\label{a:chareq}

(\ref{char-poly-Y}) can be seen as follows:
On $D_{\la}$, the quantum matrices $M^i_j$ become the operators
\beq
(\pi^i_j \tens \pi_\la)(\cR_{21} \cR_{12})
\label{RR-rep}
\eeq
acting on $V_\la$. As above, the \rep of $\cR_{21} \cR_{12}$ acting
on $V_N \tens V_{\la}$ has eigenvalues 
$\{q^{c_{\mu} - c_{\la} - c_{\la_N}}$
$=q^{2(\la + \rho)\cdot\nu -2 \la_N\cdot\rho}\}$ 
on $V_{\mu}$ in the decomposition 
\refeq{laLa-decomp}. Here $\mu = \la + \nu$ for $\nu \in V_{N}$,
and $\la_N$ is the highest weight of $V_N$.
This proves (\ref{char-poly-Y}). Note that 
if $\la$ is on the boundary of the fundamental Weyl chamber,
not all of these $\nu$ actually occur in the decomposition;
nevertheless, the characteristic equation holds.

\end{appendix}

\end{document}